\documentclass[sn-mathphys,Numbered]{sn-jnl}
\usepackage{graphicx}%
\usepackage{multirow}%
\usepackage{amsmath,amssymb,amsfonts}%
\usepackage{mathrsfs}%
\usepackage[title]{appendix}%
\usepackage{xcolor}%
\usepackage{textcomp}%
\usepackage{manyfoot}%
\usepackage{booktabs}%
\usepackage{algorithm}%
\usepackage{algorithmicx}%
\usepackage{algpseudocode}%
\usepackage{listings}%
\usepackage{xfrac}
\usepackage{comment}
\usepackage{pdfpages}
\usepackage{longtable}

\usepackage{threeparttable}

\begin{document}

\title{A snapshot review on soft-materials assembly design utilizing machine learning methods}

\author[1]{\fnm{Maya M.} \sur{Martirossyan}}\email{mmm457@cornell.edu}

\author[1]{\fnm{Hongjin} \sur{Du}}\email{hd329@cornell.edu}

\author*[1]{\fnm{Julia} \sur{Dshemuchadse}}\email{jd732@cornell.edu}

\author*[2]{\fnm{Chrisy Xiyu} \sur{Du}}\email{xiyudu@hawaii.edu}

\affil[1]{\orgdiv{Department of Materials Science and Engineering}, \orgname{Cornell University}, \orgaddress{\city{Ithaca}, \postcode{14853}, \state{NY}, \country{USA}}}

\affil[2]{\orgdiv{Department of Mechanical Engineering}, \orgname{University of Hawai`i at Mānoa}, \orgaddress{\city{Honolulu}, \postcode{96822}, \state{HI}, \country{USA}}}

\abstract{Since the surge of data in materials science research and the advancement in machine learning methods, an increasing number of researchers are introducing machine learning techniques into the next generation of materials discovery, ranging from neural-network learned potentials to automated characterization techniques for experimental images. 
In this snapshot review, we first summarize the landscape of techniques for soft materials assembly design that do not employ machine learning or artificial intelligence and then discuss specific machine-learning and artificial-intelligence-based methods that enhance the design pipeline, such as high-throughput crystal-structure characterization and the inverse design of building blocks for materials assembly and properties. 
Additionally, we survey the landscape of current developments of scientific software, especially in the context of their compatibility with traditional molecular dynamics engines such as LAMMPS and HOOMD-blue.}


\maketitle

\section{Introduction}\label{sec1}

\begin{figure}[ht]
    \centering
    \includegraphics[width=\textwidth]{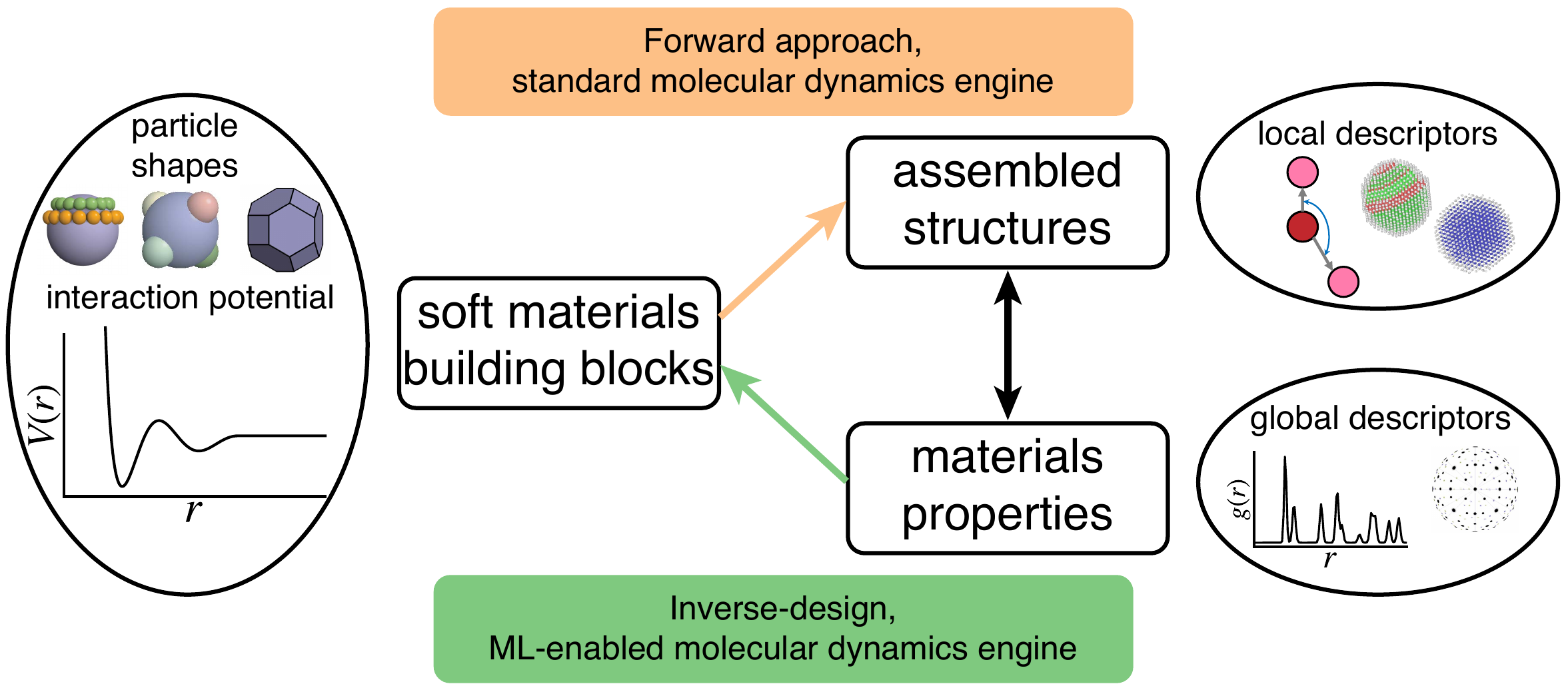}
    \caption{\textbf{Soft materials design pipeline.} Input parameters for building blocks can be patchy particles, sphere unions, and polyhedral shapes with any arbitrary pair potential functions. To quantify materials structures and properties, a variety of descriptors can be used. Here we depict bond-order parameters and OVITO's adaptive-CNA for local descriptors, radial distribution functions and bond-orientational order diagrams (BOODs) as examples for global descriptors. }
    \label{fig:1}
\end{figure}

The design of soft materials assemblies with targeted structures and properties requires the engineering of building blocks and interactions that can spontaneously assemble a target material. Before the upsurge of computational capabilities, many studies of soft materials assemblies followed a similar framework: identify a few parameters (building block properties, densities, etc.), run forward simulations varying the parameters, outline phase diagrams based on these parameters, and iterate. 
This ``forward approach'' has provided researchers with valuable insights and tools for exploring soft materials systems: phase diagrams for systems of hard spheres, anisotropic particles with polyhedral shapes, and block copolymers; rare-event sampling techniques; and local bond-order parameters to identify crystal motifs and structures. 
In recent decades, the exponential growth of computational power has widened the parameter space that can feasibly be searched, and researchers are incorporating machine learning and artificial intelligence (ML/AI) techniques to enhance their materials assembly pipelines (Fig.~\ref{fig:1}). 
Not only do these advanced tools enable us to more thoroughly probe broad questions and challenges in the field---for example, the competing nature of enthalpy and entropy in determining structure formation, dynamics, and materials properties in physical systems---but they also allow for the pursuit of reverse- or inverse-design approaches enabled by numerical optimization. Moreover, the study of soft materials (i.e., composed of mesoscopic building blocks, e.g., nanoparticles, colloids, or block copolymers) serves as a coarse-grained version of nano- or atomic-scale phenomena and can aid in understanding how to manipulate and design significantly more complicated building blocks (e.g., macromolecules, such as proteins). 

Many prior review articles provide an overview of different ML/AI techniques that have been applied in soft materials design, such as active and transfer learning \cite{ferguson_data-driven_2022} or neural networks for structural representation \cite{clegg_characterising_2021} and property design \cite{kadulkar_machine_2022}. These reviews focus heavily on novel ML algorithms and their application to soft matter. 
By contrast, this snapshot review will discuss the physical inspiration and insights that can be gleaned from adding ML/AI approaches to the quest for designing self-assembled soft materials. Given the modular nature of the soft-matter design pipeline, various ML/AI strategies can be applied to different stages of the process, and a combination of ML/AI and ML/AI-free strategies can be used to strike a balance between high predictive power and limited computational resources. 

\begin{figure}[ht]
    \centering
    \includegraphics[width=\textwidth]{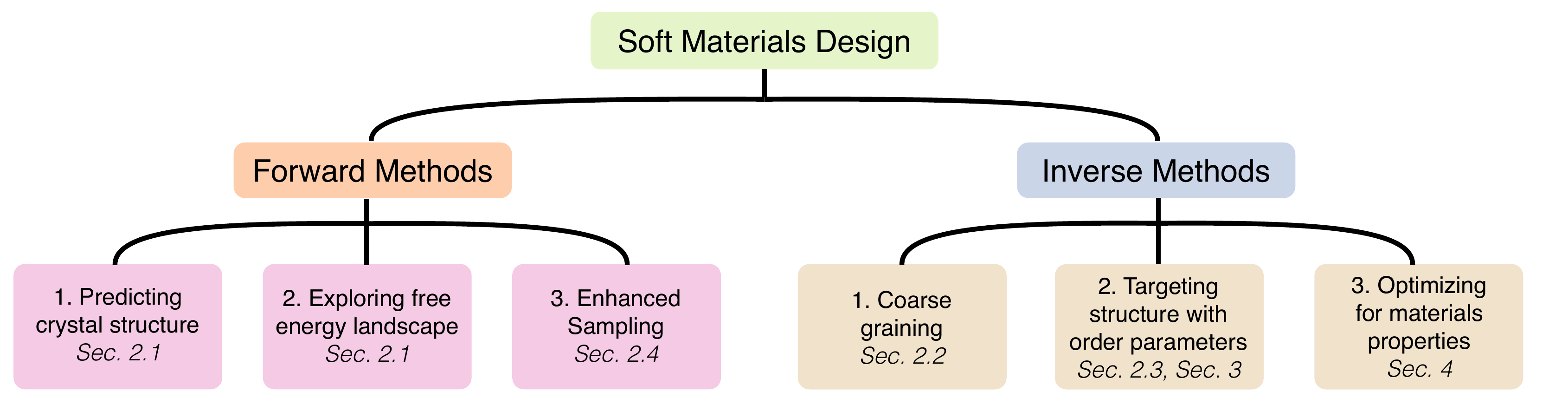}
    \caption{Overview of different forward and inverse methods for soft-materials design.}
    \label{fig:2}
\end{figure}

Firstly, we discuss the current state of ML-free techniques developed over many decades to study soft matter in both simulation and experiment. Secondly, we describe ML/AI-aided methods for different facets of materials assembly (also shown in Fig.~\ref{fig:2}): novel descriptors for quantifying local or global structure, an inverse-design framework aided by automatic differentiation, and materials property design aided by ML/AI. Lastly, we discuss the capabilities of various molecular-dynamics (MD) engines in incorporating ML tools and summarize existing ML-based descriptors by their software, methods, and their accessibility to researchers based on the computing resources needed. We intend to elucidate the state of the available methods in the field, give context for the development of the plethora of new tools created in the last few decades, and chart out how we can use these in the study of soft materials design in the future.

\begin{comment}
related reviews / perspectives: 
\begin{itemize}
\item \cite{jackson_recent_2019}
\item \cite{ferguson_machine_2018}
\item \cite{kadulkar_machine_2022}
\item \cite{sherman_inverse_2020}
\item \cite{billinge_machine_2024} 
\item \cite{clegg_characterising_2021} 
\item \cite{bedolla_machine_2021} 
\item \cite{barrat_soft_2024}
\end{itemize}
\end{comment}

\section{ML/AI-free materials design}
\subsection{Crystal structure prediction}
In computer simulations of soft materials assembly, an approximation of interparticle interactions is created and employed to predict the structure and properties of the materials system. In such \emph{forward} approaches, building blocks and interactions may also be tuned experimentally.  While interactions among all components in a system can be well-defined, \emph{a-priori} knowledge of the stable or metastable crystal structures that form is not straightforwardly obtained. Crystal structure prediction stands as one of the central challenges in materials systems and is necessary for controlling polymorphism in, for example, pharmaceutical research \cite{neumann_combined_2015}. 
Here we discuss general structure-prediction methods used in modeling---not only of atomistic systems---but in particular of soft matter systems.

Systems explore their free-energy landscape through dynamics, yet despite the ergodic hypothesis, molecular simulations may not be able to access the entirety of their phase space within a finite time frame. Several methods---simulated annealing, genetic algorithms, and enhanced sampling---have been utilized to answer questions about the global minimum (i.e., stable) structures that may be difficult to access via computational methods. 

Simulated annealing \cite{kirkpatrick_optimization_1983}---derived from the analogy to physical annealing---is a computational technique that aims to locate the global minimum of a cost (or energy) function and was developed as one of the earliest global optimization techniques. This is achieved through a gradual cooling that leads the system from an initially random configuration to an equilibrium crystal structure. An example of its application to soft materials is the prediction of binary crystal structures of oppositely charged spherical colloids \cite{hynninen_prediction_2006}. 

The Monte-Carlo-based basin hopping method \cite{wales_global_1997,wales_global_1999} explores rugged energy landscapes by hopping among the local minima (i.e., basins) using a Metropolis criterion, and it has been employed to determine the global energy minimum of size-selected clusters in two distinct hierarchical self-assemblies of triblock patchy particles \cite{morphew_programming_2018}. 

The genetic algorithm used for atomistic structure prediction \cite{oganov_crystal_2006,lonie_xtalopt_2011} mimics concepts from Darwinian evolution and selects an optimal structure from a set of candidates through a process akin to procreation: structural features from pairs of candidate structures are combined through a crossover algorithm, and new features are introduced to individual structures with a mutation algorithm. Eventually good features are preserved during `procreation' through a defined cost function. Genetic algorithms have also been used to predict stable candidates structures of patchy particles \cite{bianchi_predicting_2012} and DNA-grafted particles \cite{srinivasan_designing_2013}.
There are many other global optimization algorithms, such as metadynamics, particle swarm optimization, and landscape paving that we do not address here. 
 
\subsection{Coarse-grained models} \label{sec:CG_models}

Coarse-grained (CG) models are developed as reduced-resolution descriptions of a system to perform simulations on a larger time- and length-scale at the cost of fine-grained details. Upon treating groups of atoms as single CG particles, the subsequent challenge is to model interactions between these CG particles.
Generally, CG potentials can be derived by: (1) fitting parameters of given potential functions to reproduce target structures or thermodynamic properties, derived from atomistic simulations or empirical measurements; (2) calculating them from the direct interactions between the grouped atoms \cite{brini_systematic_2013}. 
Coarse graining has wide applications in studying soft matter systems (with relevant reviews included in the SI). Below we briefly review three categories of coarse-graining techniques, serving as essential conceptual foundations that underpin the development of ML-based approaches in optimization and parameterization. 

\subparagraph{Iterative Boltzmann inversion \& inverse Monte Carlo}
Both the iterative Boltzmann inversion (IBI) \cite{reith_deriving_2003} and inverse Monte Carlo (IMC) \cite{lyubartsev_calculation_1995} methods use a figure of merit computed directly from the structure to iteratively refine the free energy surface of the system. 
The radial distribution function (RDF) of pairwise interparticle distances is a common method in materials science for fingerprinting a crystal structure 
and can serve as a figure of merit for both IBI and IMC. 
IBI iteratively refines the potential of mean force (PMF) using Boltzmann inversion until the RDF measured in the system converges to that of the target structure. 
IMC (or reverse Monte Carlo---RMC) is an iterative procedure that is very similar to IBI, but derives pair potentials differently during the iteration using an exact update scheme with the Jacobian matrix of the RDF with respect to the potential, instead of the empirical update scheme used in IBI. 
Since IMC takes into account correlations of observables in multicomponent systems, it has a higher computational cost than IBI which can lead to convergence problems. Detailed comparisons of these two methods can be explored further in \cite{rosenberger_comparison_2016}. 
Note that the Henderson theorem states that only one pair potential is uniquely determined by a given RDF under given conditions of temperature and density \cite{henderson_uniqueness_1974}, yet the accuracy required to distinguish RDFs produced by two different pair potentials is beyond what is needed in practical use. Therefore, additional thermodynamic properties (such as pressure \cite{muller-plathe_coarse-graining_2002}) can be integrated into the optimization process alongside the RDF. 

\subparagraph{Force matching \& multi-scale coarse graining}
In contrast to the aforementioned structure-based methods (IBI, IMC), the force matching (FM) method does not aim at reproducing target distributions of structural descriptors such as the RDF. Instead it fits potentials by minimizing the difference between the CG forces and the forces in the underlying fine-grained system \cite{ercolessi_interatomic_1994}. The parameterization of the CG model is realized in a non-iterative way: the force of each atom in a CG particle is taken into account in calculating the force on that CG particle, and the minimization of force difference can be described as a least-squares problem given a sufficiently large number of snapshots (i.e., configurations) from the atomistic trajectory. 
Force matching was further extended to the multiscale coarse graining method, wherein the multibody potential of mean force is approximated by deriving effective pair potentials directly from the underlying atomistic potentials \cite{izvekov_multiscale_2005}. 

\subparagraph{Relative entropy}
The relative entropy $S_{\text{rel}}$---also known as the Kullback--Leibler (KL) divergence---is adopted from information theory and is a type of statistical distance that measures the disparity---or \emph{relative entropy}---between two probability distributions.
For coarse graining, $S_{\text{rel}}$ measures the information loss using the probability density distributions of atomistic ($P_\text{A}$) and CG models ($P_\text{CG}$):
$S_{\text{rel}} = \sum_{i} P_\text{A} \ln{\frac{P_\text{A}(i)}{P_\text{CG}(i)}}$, where $P(i)$ is the probability of configuration $i$ in a given ensemble.
The minimization of the relative entropy has been applied to the quantification of phase-space overlap between two molecular ensembles \cite{wu_phase-space_2005}, CG model development \cite{shell_relative_2008,chaimovich_coarse-graining_2011}, calculation of free-energy differences \cite{bilionis_free_2012}, and inverse design of isotropic interactions that promote self-assembly of structures including multi-component crystals \cite{pineros_inverse_2018} and colloidal strings \cite{banerjee_assembly_2019}. 
The relative entropy formalism is connected to other coarse-graining approaches insofar as they can lead to the same results depending on how potentials are modeled \cite{chaimovich_coarse-graining_2011}. While IBI and IMC are limited to optimizing pair potentials, relative entropy provides a more general framework for handling many-body CG potentials \cite{brini_systematic_2013}.

\subsection{Inverse Design}
Coarse graining and inverse methods share the goal of identifying a set of parameters of a model that best reproduces the target distribution. In fact, we can view the development of CG models as solving an inverse design problem where the target properties are the forces from the respective fine-grained systems. 
Furthermore, both coarse graining and inverse methods are fundamentally rooted in the pursuit of a more systematic framework for materials design and discovery.

A multitude of inverse methods for soft-matter self-assembly and design have been discussed in a recent review \cite{sherman_inverse_2020}. In particular, here we highlight the methods used in the inverse design of isotropic pair potentials that define short-ranged forces only by interparticle distance. Counterintuitively, the simplicity of these interactions does not compromise the structural diversity exhibited by systems that interact with such forces \cite{dshemuchadse_moving_2021}, and they can provide insight into the underlying mechanisms of self-assembly. Isotropic interactions are experimentally realizable by tuning, for example, the interactions of the isotropic DNA shell of functionalized nanoparticles \cite{mao_regulating_2023}. 

The concept of tailoring potentials to maximize the difference in the ground-state energy between the target structure and its competitors has been successfully applied to the inverse design of structures in multiple systems, including the square and honeycomb lattices in 2D \cite{rechtsman_designed_2006,marcotte_optimized_2011}, and simple cubic \cite{rechtsman_self-assembly_2006}, diamond \cite{jain_inverse_2013}, and wurtzite structures \cite{rechtsman_synthetic_2007}.

Relative entropy minimization \cite{shell_relative_2008,chaimovich_coarse-graining_2011} has also been used as a design principle for isotropic pair potentials to control the formation of pores for the assembly of porous mesophases \cite{lindquist_interactions_2017}, and to promote self-assembly of 2D and 3D crystals \cite{lindquist_communication_2016, pineros_inverse_2018}, colloidal strings \cite{banerjee_assembly_2019}, as well as size-specific cluster fluids \cite{lindquist_inverse_2021}. This ``on-the-fly'' approach uses structures generated during each optimization step of the particle interactions, 
thereby promoting the self-assembly of the target structure from a disordered state. This optimization process was also employed in combination with Fourier-space filters to design simple interactions that could be more experimentally feasible \cite{adorf_inverse_2018}. 

All these approaches to modifying interactions or building blocks can be encompassed by ``digital alchemy,'' which was first introduced as a statistical-thermodynamics method to inversely design anisotropic particle shapes that favor the self-assembly of a target structure with Monte Carlo simulations \cite{van_anders_digital_2015}. The general framework of describing particle attributes as thermodynamic variables---allowing them to fluctuate, and as a result identifying attributes crucial for controlling self-assembly---has also been extended to MD simulations with success for a handful of structures \cite{zhou_alchemical_2019,zhou_inverse_2021}. 

\subsection{Enhanced sampling}
Enhanced sampling encompasses a class of methods that enables the simulation of hard-to-reach states. There are many different flavors of enhanced-sampling methods: umbrella sampling, replica exchange, metadynamics, and simulated annealing to name a few. Most enhanced-sampling methods apply a bias force or potential to drive the system to explore the region of phase space containing states of interest. These states are often described by a set of collective variables (also referred to reaction coordinates, order parameters, or structural descriptors in other contexts). We refer the reader to the SI for many in-depth reviews on different aspects of enhanced sampling.

\section{Descriptors for self-assembly studies}\label{sec:descriptors}

There is a rich history of using order parameters to define and study phase transitions in physical systems, allowing for the most important variables or degrees of freedom to be captured. Reducing a physical system's $3N$ spatial dimensions to a more ``natural'' low-dimensional representation extracts the \emph{most relevant} characteristics of the system's behavior. In the study of self-assembly and growth, order parameters---i.e., structural descriptors---vary widely in their physical basis and in the behavior of interest of the physical system for whose study they are being used. 

Specific variable choices are often necessary to define an order metric, but they can also prove limiting or insufficiently descriptive when confronted with an increasing variety of behaviors or motifs in a single system. For example: how can a descriptor be designed to study a growing crystal with multiple crystalline environments, each with a different kind of crystalline symmetry? 

Here, we highlight conventional approaches using physically-inspired descriptors and how coupling these methods with machine learning techniques---well-suited for leveraging and interpreting high-dimensional data---allows for a more complete picture of self-assembly to emerge across a variety of physical systems. In our discussion, we place significant emphasis on the physical basis of descriptors rather than on the specific ML tools utilized, in part because these physical descriptors should be tailored to the given system or behavior being studied, and in part because of evidence suggesting that the optimization schema used does not significantly change the outcome of an ML-based analysis approach \cite{mao_training_2023}. Later, in Section~\ref{sec:descriptor_table}, we highlight the technical ML specifics for many of the methods discussed in this section.

Local descriptors that accurately quantify structural motifs have been used to develop atomistic machine-learned interatomic potentials (MLIAPs) and led to marked improvements over potentials calculated from electronic structure alone \cite{behler_perspective_2016}. The difference in applying descriptors to soft matter is the lack of atomic or energetic data to train on, in order to predict resultant properties; this, in part, explains why solving the inverse-design question represents such a ``holy grail'' for the field. Consequently, good structural descriptors are critical for capturing and optimizing system behavior. 

As is true for MLIAPs, a good descriptor should be immutable upon equivalent configurations generated by translations, rotations, and permutations to a motif or crystal structure. This mathematical property with respect to a symmetry operation is referred to as invariance (equivariance or covariance also satisfy the required criteria). While not required, differentiability is particularly useful for applications utilizing automatic differentiation methods such as JAX-MD \cite{schoenholz_jax_2021} (see Sec.~\ref{sec:jax-md} for a detailed description).

\begin{figure}[ht]
    \centering
    \includegraphics[width=0.8\textwidth]{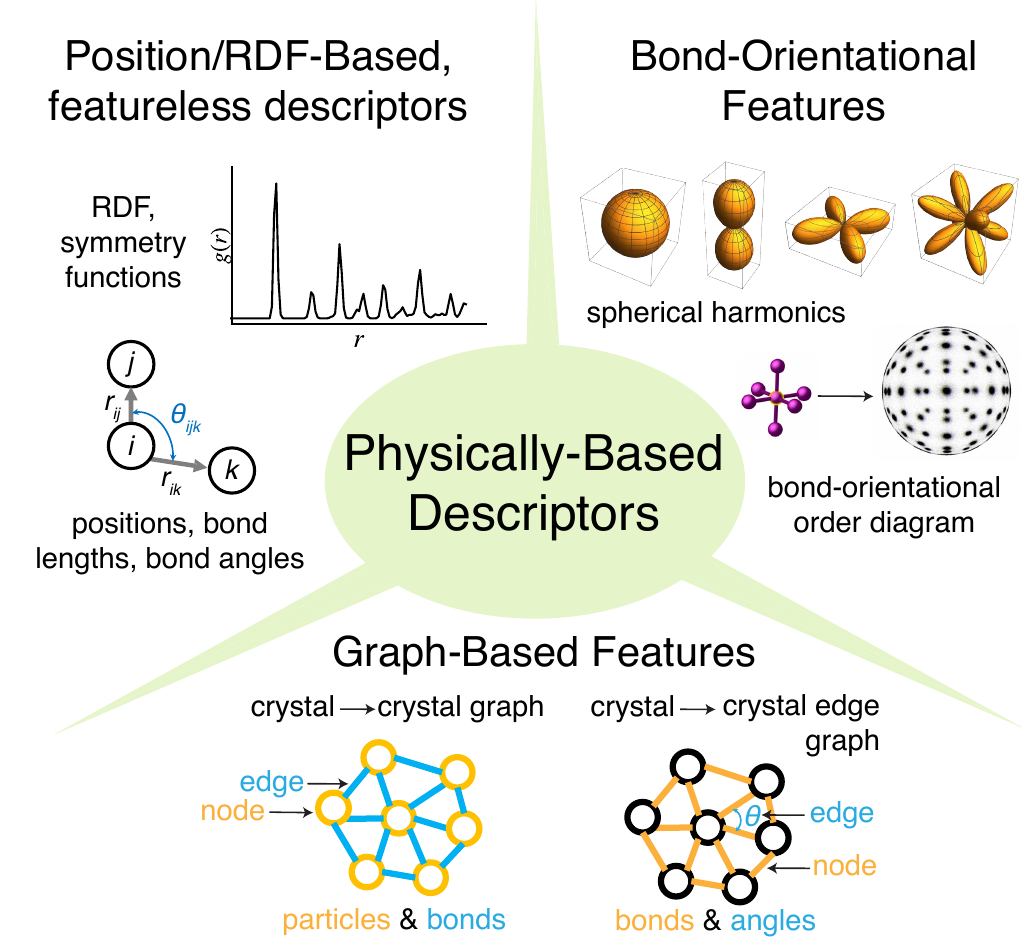}
    \caption{The three broad classes of descriptors: position-/RDF-based and featureless descriptors, bond-orientational features, and graph-based features.}
    \label{fig:3}
\end{figure}

We discuss several structural descriptors, broadly grouped according to their physical basis (as shown in Fig.~\ref{fig:3}): (1) structure-only parameters including RDF-based, position-based, as well as ``descriptor-free'' (featureless) parameters that are exclusively machine-learned; (2) bond-orientational features utilizing spherical harmonics in a variety of approaches; (3) graph-based or topological features. We aim to provide a comprehensive overview of the featurizations in the field, although inevitably we will be unable to cover all relevant work in the scope of this snapshot review. 

We include in the SI an additional summary on thermodynamics-inspired features and methods, although these are less commonly used and not as effective compared to structure-based approaches. We will sidestep a common problem lurking among many of the discussed methods, which is exactly how neighbors or radial cutoffs are chosen---handled differently by each method. Finally, we largely ignore informatics approaches (for example, the Polymer Genome platform \cite{kim_polymer_2018}) as they use hundreds of descriptors in a hierarchical manner to train models that target properties. This section will focus on work that uses a specific choice of descriptor and its appropriate use cases.

\subparagraph{RDF-based and position-based features}

Using positional data with minimal manipulation is a logical choice for a structural descriptor. The oldest criterion for melting is the Lindemann order parameter \cite{lindemann_uber_1910}---based on particle mean-squared displacement from equilibrium position---utilized in soft matter studies of nucleation and growth dynamics \cite{peng_two-step_2015}. In hard-disk or sphere systems, phase transitions were commonly detected using only sorted neighbor distances (i.e., in the first shell of the RDF) as features with unsupervised learning \cite{jadrich_unsupervised_2018}. 

\citet{behler_generalized_2007} introduced radial and angular symmetry functions---representing potential-energy surfaces in atomic systems---that also bear similarity to the RDF but are localized to a particle's environment. Such symmetry functions have been utilized in the development of ML-based structural identification methods for complex phases in polymorphic systems such as ice \cite{geiger_neural_2013} or the ML-based order metric ``softness'' for identifying particle susceptibility to rearrangements \cite{cubuk_identifying_2015}. Softness has been used to predict glass dynamics \cite{schoenholz_structural_2016} as well as to identify grain boundaries \cite{sharp_machine_2018} and improve growth models \cite{freitas_uncovering_2020} in atomistic MD simulations. 
Other position-based features include using bond angle, bond length, and interparticle separation distance as inputs for an unsupervised crystal-structure identification method \cite{reinhart_unsupervised_2021}, defining a loss function based on a ``stencil'' used to target assembly of a specific polymorph using JAX-MD \cite{goodrich_designing_2021}, or utilizing particle positions and particle-level features to build geometric algebra-based representations of structure with deep learning \cite{spellings_geometric_2022}. 

\subparagraph{Featureless order parameters} 

``Descriptor-free'' or featureless order parameters can be conceptualized as a subcategory of position-based features, but they differ in that they use entirely unmanipulated data that must be interpreted using statistical or machine-learned methods. Because these descriptors do not use representations that are invariant to translations, rotations, and permutations, they instead rely on data augmentation---that is, the model must learn these symmetries from an abundance of data in a variety of configurations, rather than invariance being built into the inputs for training. 

Featureless unsupervised learning methods have so far been used to identify magnetization phase transitions in the canonical two-dimensional Ising model \cite{wang_discovering_2016, wetzel_unsupervised_2017} using entire Ising spin matrices as inputs. Unlike in the Ising model, a ``descriptor-free'' approach is more difficult to apply to systems where particle positions are variable, but this has been accomplished \cite{defever_generalized_2019, wang_descriptor-free_2022} relying on sophisticated model architectures such as PointNet \cite{qi_pointnet_2017} in order to perform feature extraction.
Other frameworks that use deep learning approaches with particle positions \cite{swanson_deep_2020} (or, combined with atomic-level features as inputs \cite{schutt_SchNet_2018}) could be extended to target properties in soft-matter systems.

\subparagraph{Bond-orientational features} 

Bond-orientational features differ from those described above in that they enforce spherical symmetry in their representations of local structure.
For two-dimensional structures, the $\Psi_n$ order parameter is defined by the expectation of $n$-fold symmetry in the crystalline phase, and it has been utilized in the study of colloidal crystallization experiments \cite{gasser_real-space_2001}. The Steinhardt $Q_l$ order parameter \cite{steinhardt_bond-orientational_1983} and its neighbor-averaged version $\overline{Q}_l$ \cite{lechner_accurate_2008}---which are rotationally-invariant representations of a particle's neighborhood using summations of spherical harmonics---have been used to identify local motifs or differentiate phases of matter, distinguish between simple sphere packings (\emph{bcc}, \emph{ccp}, \emph{hcp}) \cite{du_shape-driven_2017} and study quasicrystal growth \cite{keys_how_2007} in simulations of three-dimensional systems. 
Steinhardt's $\hat{W}_l$ parameter has also been utilized to identify motifs in computational studies of pre-crystallization fluids \cite{hu_revealing_2022}.

The addition of machine-learning methods to these bond-orientational approaches has allowed for the extension of order parameters to more complex crystal structures, polydisperse packings, and non-close-packed local environments in crystalline solids (i.e., expanding beyond icosahedral, \textit{fcc}, \textit{hcp}, or \textit{bcc} local environments). Spherical harmonics-based descriptors have been used with unsupervised learning approaches to distinguish highly similar, complex structures \cite{spellings_machine_2018}, as well as to distinguish between local environments and phases during the self-assembly of structures with one or more crystalline motifs \cite{adorf_analysis_2020, martirossyan_local_2024}. Supervised approaches using spherical harmonics \cite{spellings_machine_2018} or Steinhardt-based features have also successfully identified crystalline motifs in binary systems \cite{coli_artificial_2021}, and unsupervised approaches have similarly been employed with Steinhardt-based features to study local order in glasses and liquids \cite{boattini_autonomously_2020} or at crystalline grain boundaries and binary systems \cite{boattini_unsupervised_2019}.

\begin{table*}[ht]
% \begin{minipage}{\textwidth}

\caption{ML-based descriptors for materials assembly and design. Acronyms used are: support vector machine (SVM), principal component analysis (PCA), density-based spatial clustering of applications with noise (DBSCAN), Gaussian mixture model (GMM), artificial neural network (ANN), convolutional neural network (CNN), graph neural network (GNN). 
%Software listed is not necessarily comprehensive. 
Compute resources reflect those used or reported by the respective authors.}
\label{tab:descriptors}
\centering
%\begin{tabular}{c|c|c|c|c}
\begin{threeparttable}
\begin{tiny}
\begin{tabular}{p{0.95in}|p{1in}|p{0.95in}|p{0.95in}|p{0.7in}}
% \begin{tabularx}{\textwidth}{>{\small}c|>{\tiny}X|>{\tiny}X|>{\tiny}X|>{\tiny}c}
%\hline

\centering{\textbf{\small{Reference}}} & \centering{\small{\textbf{Features}}} & 
\centering{\small{\textbf{Models}}} & 
\centering{\small{\textbf{Software}}} & \small{\textbf{Compute}} \\ \hline \hline

\citet{geiger_neural_2013}
& symmetry functions
& ANN
& -
& GPU \\

\citet{cubuk_identifying_2015} 
& symmetry functions 
& SVM 
& LIBSVM\tnote{1} %\cite{chang_libsvm_2011} 
& - %probably CPU-intensive/GPU
\\ %\hline

\citet{wang_discovering_2016} 
& Ising spin matrix
& PCA\texttt{+}k-means clustering 
& -
& - % probably CPU/GPU
\\ %\hline

\citet{wetzel_unsupervised_2017} 
& Ising spin matrix 
& kernel PCA/ DBSCAN/ variational autoencoder 
& -
& - %probably CPU/GPU
\\

\citet{jadrich_unsupervised_2018}
& sorted neighbor distances
& Incremental PCA
& Sklearn
& - % probably CPU/GPU
\\

\citet{reinhart_unsupervised_2021} 
& neighbor distances, bond angles \& lengths, particle-level features 
& UMAP\texttt{+}Random Forest Classifier
& UMAP, Sklearn
& - %probably CPU/GPU
\\
% differentiable

\citet{defever_generalized_2019} 
& particle positions
& PointNet\tnote{2} % not sure if this should be moved to next column
& TensorFlow
& - % probably CPU-intensive/GPU
\\

\citet{wang_descriptor-free_2022} 
& particle positions
& autoencoder\texttt{+}GMM 
& TensorFlow, Sklearn
& - % probably CPU-intensive/GPU 
\\ %\hline

\citet{schutt_SchNet_2018} %(SchNet) 
& atomic nuclear charges \& positions
& filter-generating network 
& TensorFlow, SchNet\tnote{3} 
& CPU-intensive / GPU 
\\ %\hline

\citet{swanson_deep_2020}
& particle positions
& CNN / message-passing neural network
& TensorFlow/PyTorch, ``glassML''\tnote{4}
& GPU \\

\citet{spellings_geometric_2022} %(GAlA) 
& multivectors (geometric products of particle positions) \& particle-level features
& attention mechanism 
& Keras, TensorFlow, GAlA%\cite{spellings_klarhgeometric_algebra_attention_2023} 
\tnote{5} 
& GPU \\

\citet{spellings_machine_2018} 
& spherical harmonics (pythia)%\cite{noauthor_pythia_2022}
\tnote{6}
%\footnotemark{})
& PCA\texttt{+}GMM / ANN 
& Sklearn/ Keras  
& CPU % could be CPU/GPU 
\\ %\hline

\citet{adorf_analysis_2020} 
& bispectrum spherical harmonics (pythia)\tnote{6}
& PCA\texttt{+}UMAP \texttt{+}HDBSCAN$^{*}$ 
& Sklearn, UMAP, HDBSCAN$^{*}$
& - %likely CPU/GPU
\\ %\hline

\citet{boattini_unsupervised_2019} 
& Steinhardt parameters
& autoencoder\texttt{+}GMM 
& Sklearn 
& - \\ %probably CPU/GPU

\citet{coli_artificial_2021} 
& Steinhardt parameters
& ANN 
& Keras, TensorFlow 
& - % probably CPU/GPU
\\ %\hline

\citet{grisafi_symmetry-adapted_2018} %(SA-GPR) 
& SOAP\tnote{7}
& Gaussian process 
& SciPy, %SymPy, ASE 
%\cite{noauthor_lab-cosmosa-gpr_2024} 
SA-GPR\tnote{8} 
& CPU-intensive
\\ %hline

\citet{gardin_classifying_2022} 
& SOAP\tnote{7}
& PCA\texttt{+}PAMM/ Hierarchical clustering 
& Sklearn, PAMM%\cite{noauthor_lab-cosmopamm_2023}
\tnote{9} 
& - % probably CPU/GPU 
\\ %hline

\citet{geiger_e3nn_2022} %(e3nn) 
& irreps (tensor products of spherical harmonics) 
& CNN 
& JAX/PyTorch, e3nn%\cite{noauthor_euclidean_nodate} 
\tnote{10} 
& GPU %probably also CPU-intensive
\\ 

\citet{duvenaud_convolutional_2015}
& molecular graphs
& CNN
& SciPy, Autograd, ``Neural fingerprint''\tnote{11}
& - %probably CPU-intensive/GPU?
\\ 

\citet{bapst_unveiling_2020} 
& crystal graphs
%\cite{noauthor_deepmind-researchglassy_dynamics_nodate} 
& GNN 
& TensorFlow/TF-Replicator, JAX, ``Glassy dynamics''\tnote{12} 
& - % probably CPU-intensive/GPU 
\\ %hline

\citet{chapman_quantifying_2023} 
& crystal graphs 
& GNN 
& PyTorch%, PyTorch-Geometric
, SODAS/graphite%\cite{noauthor_llnlgraphite_nodate}
\tnote{13}
& GPU 
\\

\citet{choudhary_atomistic_2021}
& crystal/line graphs\texttt{+}radial basis functions
& message-passing GNN
& ALIGNN%\cite{noauthor_usnistgovalignn_2024}
\tnote{14}
& GPU
\\

\citet{aroboto_universal_2023} %(SODAS\texttt{++}) 
& ALIGNN%\cite{noauthor_usnistgovalignn_2024}
\tnote{14}
& UMAP\texttt{+}GNN 
& UMAP, PyTorch, %PyTorch-Geometric, Networkx, SciPy, NumPy, ASE
%\cite{noauthor_materials-informatics-laboratorysodas_nodate} 
SODAS++\tnote{15} 
& - %probably CPU-intensive/CPU
\\

\citet{reinhart_machine_2017}
& CNA-based crystal graph
& Diffusion maps 
& Neighborhood Graph Analysis (NGA)
& - %probably CPU?
\\ 

\citet{xie_crystal_2018} %(CGCNN) 
& atom-level features \& crystal graphs
& CNN
& Sklearn, PyTorch, CGCNN\tnote{16} 
& - %probably CPU-intensive/GPU?
\\

\citet{banik_cegann_2023} %(CEGANN) 
& crystal edge graphs
& attention mechanism 
& PyTorch, Sklearn, %PyTorch-Ignite, Pymatgen, Pydantic, Natsort 
%\cite{banik_sbanik2cegann_2024} 
CEGANN\tnote{17} 
& GPU \\ %\hline

\citet{sheriff_quantifying_2023} 
& crystal graphs \& particle-level features
& - %only notes use of KL-div, doesn't specify type of model (maybe not needed because it uses e3nn)
& e3nn\tnote{10} 
& - %probably CPU-intensive/GPU
\\

%\citet{hansen_machine_2015} % Bag of Bonds
%& bond energies 
%& kernel ridge regression
%& - (see chemreps)\tnote{18} %found this on github which cites the paper, but isn't necessarily made by these authors
%& CPU %from \cite{hansen_assessment_2013}
%\\ 

% \end{tabularx}
%\hline
\end{tabular}
\smallskip%\footnotesize
\begin{flushleft}
% \begin{tablenotes}
% \item[1] ...
$^1$~\texttt{https://github.com/cjlin1/libsvm}
$^2$~\texttt{https://github.com/charlesq34/pointnet}
$^3$~\texttt{https://github.com/atomistic-machine-learning/SchNet}
$^4$~\texttt{https://github.com/ks8/glassML}
$^5$~\texttt{https://github.com/klarh/geometric\_algebra\_attention}
$^6$~\texttt{https://github.com/glotzerlab/pythia}
$^7$~\texttt{https://singroup.github.io/dscribe/latest/}
$^8$~\texttt{https://github.com/lab-cosmo/SA-GPR}
$^9$~\texttt{https://github.com/lab-cosmo/pamm}
$^{10}$~\texttt{https://github.com/e3nn}
$^{11}$~\texttt{https://github.com/HIPS/neural-fingerprint}
$^{12}$~\texttt{https://github.com/google-deepmind/deepmind-research/tree/master/glassy\_dynamics}
$^{13}$~\texttt{https://github.com/LLNL/graphite}
$^{14}$~\texttt{https://github.com/usnistgov/alignn}
$^{15}$~\texttt{https://github.com/Materials-Informatics-Laboratory/SODAS}
$^{16}$~\texttt{https://github.com/txie-93/cgcnn}
$^{17}$~\texttt{https://github.com/sbanik2/CEGANN}
%$^{18}$~\texttt{https://github.com/chemreps/chemreps}
% \end{tablenotes}
\end{flushleft}
\smallskip
\end{tiny}
\end{threeparttable}
% \end{minipage}
\end{table*}

A handful of other approaches using spherical harmonics-based descriptors have been formulated for the study of atomic materials and extended to the study of phase transitions or soft and molecular systems. The Smooth Overlap of Atomic Positions (SOAP) descriptor \cite{de_comparing_2016}---which utilizes spherical harmonics to represent Gaussian-smeared particle densities---has also been adapted for ML-based studies of materials: a Gaussian process with a SOAP kernel \cite{grisafi_symmetry-adapted_2018} or unsupervised methods with a SOAP descriptor
\cite{gardin_classifying_2022} 
have been used to study the formation of (supra)molecular materials. Euclidean neural networks (e3nn) \cite{geiger_e3nn_2022} use spherical harmonics to create irreducible representations that leverage equivariance to learn symmetry-based translations and rotations, and they can be used to define order parameters that identify the breaking of these symmetries (e.g., during a phase transition) \cite{smidt_finding_2021}.

\subparagraph{Graph-based features} 

Yet another intuitive way to featurize inter-particle bonding structure are graph-based features---not to be conflated with graph neural networks, although they can appear together. Graph-based features include particle connections, bond lengths and angles, and local neighborhood geometry in their representation of local structure. The most popular graph-based feature is referred to as Common Neighbor Analysis (CNA) \cite{honeycutt_molecular_1987, faken_systematic_1994}, a tool which classifies simple 3D motifs by the topology of particle neighborhoods and which is integrated (along with its variants) into the ``Open Visualization Tool'' (OVITO)\footnote{\texttt{https://www.ovito.org}}. CNA has been applied to numerous studies of crystallization, such as the simulation study of charge-stabilized colloidal suspensions \cite{urrutia_banuelos_common_2016}.
Another commonly used method for simple crystal-structure or motif identification is polyhedral template matching (PTM) \cite{larsen_robust_2016}, which uses the convex hull of neighbors around a particle to create a planar graph and performs template matching to identify motifs. 
Given the success of these approaches, the addition of machine-learning methods is highly sensible and allows a larger variety of local structures to be represented and identified as compared to CNA and PTM, particularly for systems where atom- or particle-level features are important (such as having two different particle sizes or components).

Graph neural networks (GNNs) are designed to take in graphs as inputs and perform convolutions to create embeddings of local structure. 
Crystals lend themselves naturally to representations as planar graphs, where nodes and edges represent particles and bonds.
For example, a GNN is used to build local descriptors from graph-based features that can identify disorder such as in grain boundaries or interfaces \cite{chapman_quantifying_2023}. 
GNNs have also been used with `crystal edge graphs' rather than crystal graphs, where nodes represent bonds in the crystal and edges represent bond pairs (i.e., angles between bonds).
Recent work has identified phase transitions by building global descriptors \cite{aroboto_universal_2023} with the Atomistic Line Graph Neural Network (ALIGNN) \cite{choudhary_atomistic_2021}, 
which uses a GNN to create latent representations using message-passing between the crystal graph (interatomic bond graph) and the crystal edge graph (line graph corresponding to bond angles). In a similar vein, crystal edge graphs have been used to perform crystal identification tasks on individual particles \cite{banik_cegann_2023}.

However, GNNs are not the only types of ML approaches utilized with graph-based features. Convolutional neural networks have been applied to graph-based features for molecules \cite{duvenaud_convolutional_2015}, atomic structures \cite{xie_crystal_2018}, and glasses \cite{bapst_unveiling_2020}. Other ML methods have also been applied with graph-based features---for example, diffusion maps for local environment identification including both amorphous and crystalline structures \cite{reinhart_machine_2017}, or for building representations of chemical ordering in multi-component alloys for use in a relative entropy-based order metric \cite{sheriff_quantifying_2023}.

\section{Designing for properties}

Frequently, the design of materials with specific properties has been handled separately from the design of assembly pathways to target particular structures \cite{sherman_inverse_2020}. 
Inverse design for the properties of a material requires the use of the property itself as the figure of merit of the computation, enabling the use of any property that can be computed from a material's structure. 
Machine learning facilitates the accelerated evaluation of complex structure--property relationships that would otherwise be prohibitive as a computational figure of merit \cite{sherman_inverse_2020}. 
This can be done by reducing the dimensionality of the order parameter or by applying supervised machine learning directly to the structure--property relationship \cite{kadulkar_machine_2022}. 

Many advances have been made in the field of atomistic (``hard'') materials such as alloys, 
and electronic materials, 
in which atomic structure can directly be related to bulk properties of a material \cite{xie_crystal_2018,choudhary_atomistic_2021}. 
In this snapshot review, we concentrate on mesoscopic (``soft'') matter. While bulk properties can often be computed directly in hard-matter systems---because the relevant structural features occur on only one length scale---soft-matter systems exhibit multiple salient length scales, making predictive modeling of a variety of properties more challenging. For length consideration, we provide our discussion on soft materials properties in the SI, where we discuss five different kinds of materials properties: (1) mechanical properties, (2) thermodynamic and phase properties, (3) electronic and optical properties, (4) transport properties, and (5) chemical properties.

\section{Software developments}

Many well-known MD packages were developed before the popularity of ML-enhanced materials research, so integrating ML methods with traditional MD simulations can prove challenging. In this section, we briefly review the compatibility of current MD engines with ML methods, how software for descriptors can be used, and we showcase a new MD engine that is intrinsically compatible with the current ML/AI software packages.

\subsection{Integration with ML methods for traditional MD engines}

Traditional MD packages such as LAMMPS\footnote{\texttt{https://www.lammps.org}} and HOOMD-blue\footnote{\texttt{https://glotzerlab.engin.umich.edu/hoomd-blue}} are extremely powerful engines that can perform MD simulations very effectively. Although primarily written in C\texttt{++}, there are now tools available to integrate ML methods with these MD platforms. 

LAMMPS hosts a well-documented webpage\footnote{\texttt{https://www.lammps.org/external.html}} providing a list of software packages that are either external---and built on top of LAMMPS---or standalone---either providing input parameters for LAMMPS or other MD engines, or incorporating LAMMPS as one of their MD engines to produce simulation trajectories. Within the list, there are packages linking PyTorch with LAMMPS and several packages for ML-based interaction potentials. 

Similarly, HOOMD-TF\footnote{\texttt{https://github.com/ur-whitelab/hoomd-tf}} was developed to link TensorFlow with HOOMD-blue (currently compatible with HOOMD-blue 2.6+ but not 3.x, etc., due to a major API change\footnote{A---to-date unmerged---branch exists on the HOOMD-TF GitHub page, allowing to make the code compatible with HOOMD-blue 3.x.}).

\subsection{Software and methods overview for descriptors} \label{sec:descriptor_table}

There is a large variety of methods for quantifying structural order that are applied to study crystal growth and assembly---which are equally as diverse as the open questions in the field. With many ML-based methods being developed for different use cases and specific physical systems, we highlight methods in Table.~\ref{tab:descriptors} with the most important software and architecture details as well as computing resources needed for each.

\subsection{JAX-MD} \label{sec:jax-md}

Computing derivatives or gradients is a crucial component of many machine-learning techniques. Utilizing general-purpose automatic differentiation \cite{baydin2018automatic} implementations is standard in many different machine-learning packages such as PyTorch, Julia, and MatLab's Deep Learning Toolbox. Similarly, various materials-science applications also require the computation of gradients ranging from force computation in MD to evaluating stress tenors for materials properties.

Following the release of JAX\footnote{\texttt{http://github.com/google/jax}} in 2018---a Python-based software package that enables end-to-end differentiation---various packages were developed utilizing JAX's new ability to differentiate through complicated functions. JAX-based materials-science software packages are not limited to JAX-MD, and include JAX-AM\footnote{\texttt{https://github.com/tianjuxue/jax-am}}, 
JAX-FEM\footnote{\texttt{https://github.com/deepmodeling/jax-fem}}, 
and GradDFT\footnote{\texttt{https://github.com/XanaduAI/GradDFT}}.  
Given the scope of this snapshot review, which concentrates on assembly design, we will only highlight work related to JAX-MD.

\begin{figure}[H]
    \centering
    \includegraphics[width=0.9\textwidth]{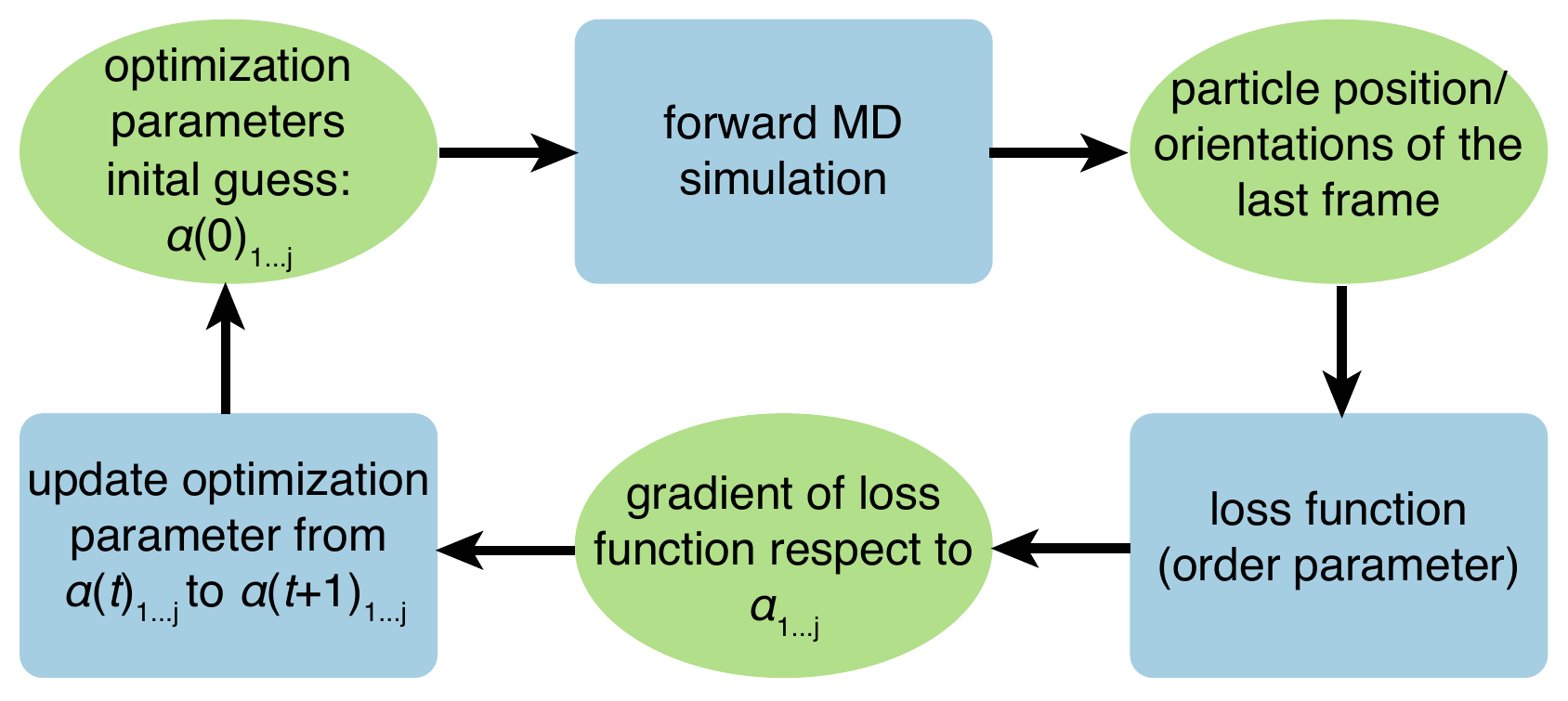}
    \caption{Inverse-design framework based on JAX-MD.}
    \label{fig:4}
\end{figure}

The molecular-dynamics engine JAX-MD currently features simulation environments to model isotropic pair potentials and anisotropic particles using rigid-body constructions with standard integrators such as NVE, NVT, NVP, Brownian dynamics, and Langevin dynamics. As JAX-MD is written fully in Python, the overhead for any user to define a new pair potential, external field---or interface with other ML/AI methods---is minimal. Moreover, when implementing a new pair potential, no additional force implementation is needed as gradients (i.e., derivatives) of the interaction potential can be retrieved directly to update quantities such as particle velocity and acceleration.

So far, JAX-MD has been used to design assembly and transition rates for colloidal systems \cite{goodrich_designing_2021}, anisotropic building blocks for bulk and finite assembly \cite{king_programmable_2023}, controlled disassembly of colloidal clusters \cite{krueger_tuning_2023}, error-free polymer growth \cite{zhu_proofreading_2023}, and minimal-work pathways in non-equilibrium systems \cite{engel_optimal_2023}. These papers showcase the breadth and versatility of the physical systems and properties that JAX-MD can model and design. Generally, the optimization regime in JAX-MD is system-agnostic as long as the user can provide a loss function / order parameter with meaningful gradients to describe the simulated system. 

Fig.~\ref{fig:4} illustrates a schema for the use of JAX-MD for inverse design, but this is not the only way to implement such a workflow. For example, one can update building-block properties after a fixed number of simulation steps instead of at the end of one round of forward simulation. Here, we want to provide a working example as a starting point for interested researchers to explore. Apart from using JAX-MD to inversely design assembly, yet another unexplored territory for JAX-MD is to combine it with enhanced-sampling methods. Computing forces for a bias potential in MD can be challenging to implement, but with the help of automatic differentiation, no explicit force implementation would be needed.\footnote{See \texttt{https://colab.research.google.com/drive/1eOBqUlRxhUvPsxfl9hGcVwHTleN2GNBJ} for an example.}

\subsection{Accessibility}

In this section, we highlight a few barriers to accessing some of the methods discussed in this snapshot review. Computing resources are vital to those who may want to train models or utilize inverse-design methods for their research. Methods that run on GPUs \emph{can} be run on CPUs as well, but the difference in wall time can amount to orders of magnitude---especially for tasks that require backpropagation. Moreover, tasks that require backpropagation or automatic differentiation can be GPU-memory intensive, sometimes requiring the most advanced GPUs with 80 GB of memory. Therefore, we encourage including computational resources either used (or utilizable) for training models and describing the associated computational costs for new methods being published. In a similar vein, sharing code on open-source platforms like GitHub is increasingly common and can function as a ``plug and play'' tool for non-experts to utilize.

Another possible barrier is the need to transmute training data (from simulation or experiment) into the expected data format for a specific method. For example, many of the approaches highlighted in Tab.~\ref{tab:descriptors} rely on specific file formats (typically only used for data output by a particular MD engine). Tools such as the \textit{garnett} software\footnote{\texttt{https://garnett.readthedocs.io}} can help with reading/writing to/from different simulation file formats, although not all common formats are included (e.g., the \texttt{.xyz} file format). A similar issue arises with ML-specific backends: often methods are developed for only one of the three---Keras/TensorFlow, JAX, or PyTorch. These are just a few of the ``language barriers'' that arise from the diversity of computing tools that researchers use.

Finally, access to large volumes of data for training models is usually straightforward for simulators, and there are already databases hosting services for more various materials datasets \footnote{\texttt{https://www.materialsdatafacility.org}}---but this is not necessarily the case for soft-matter experimentalists. While we do not necessarily endorse the publishing of trained models as a solution, we urge consideration of how models can behave for low data-volume cases. These considerations could be especially important in developing simulation--experiment pipelines for training models or inverse-design approaches.

\section{Conclusion}\label{sec4}

In this snapshot review, we discuss many approaches used in the optimization and design of soft materials such as structure prediction, coarse-graining, enhanced sampling, and how these approaches are not only compatible with but enhanced by ML methods, as well as a variety of software that can be used to target specific structures or materials properties. We discuss both forward approaches---critical for the study of phase behavior and self-assembly---as well as more targeted inverse approaches that are used specifically for design. 

Through our survey of methods in the field, we emphasize the importance of the physical basis of methods and features. We also include relevant methods that are developed for atomic systems as these approaches can be extended to soft or mesoscale materials. We hope that this snapshot review can serve as a guide for those looking to apply (or create) ML-based methods for scientific questions.

We also offer a few reflections on how we believe the methods we review can be best used going forward. Given the multitude of descriptors and inverse design tools developed in just the last decade, the fields of both atomistic and soft materials are ripe for employing new methods to conduct scientific research. That is, the ``low-hanging fruit'' of ML-based approaches are being or have been picked, and further development or use of methods should be tailored to answering open questions in the field or addressing specific design principles. Bridging the gap between the tool-makers and tool-users will be imperative in order to address open scientific questions and to connect theory, simulation, and experiments: these range from the need for robust descriptors that can handle particle-locating in experiments, to more fundamental questions such as the effect of interaction and structure in particle-based systems. We are optimistic that with the newly available avenues---provided by the power of machine learning and the multitude of new computational approaches built upon decades of progress---we can answer fundamental questions regarding structure formation and design of matter across various length scales in the future. 

\backmatter
\bibliography{main_arxiv}

\subparagraph{Funding:} This material is based upon work supported by the National Science Foundation under Grant No.\ DMR-2144094, the National Science Foundation AI Institute in Dynamic Systems under Grant No.\ CBET-2112085, as well as the Camille and Henry Dreyfus Foundation through a Machine Learning in the Chemical Sciences and Engineering Award (ML-22-038). 
M.\ M.\ M.\ acknowledges support from the National Science Foundation Graduate Research Fellowship Grant No.\ DGE-2139899 (2021--2024). 
\subparagraph{Conflict of interest:} On behalf of all authors, the corresponding author states that there is no conflict of interest.
\subparagraph{Author contributions:} M.\ M.\ M., H.\ D., J.\ D., and CX.\ D. wrote the manuscript.

\newpage

\includepdf[pages=-]{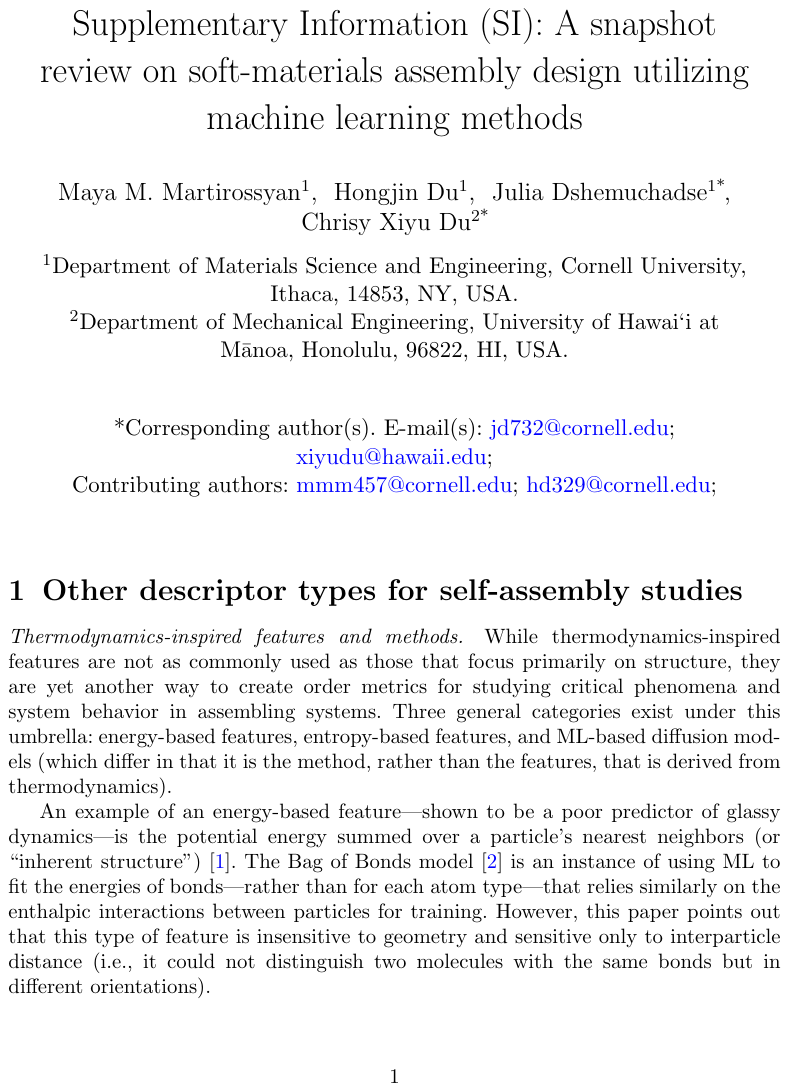}

\end{document}